 \documentclass[11pt]{article}
 
\usepackage{graphicx}
\usepackage{tabularx}
\usepackage{amsmath, amssymb, dsfont, accents}
\usepackage{listings,multicol, multirow}
\usepackage{array}
\usepackage{fullpage}
\usepackage{color}
\usepackage{siunitx}
\usepackage{booktabs} 
\usepackage[disable]{todonotes}
\usepackage{url}

\usepackage{tikz,pgfplots}
\usepackage{tikzscale}
\usetikzlibrary{arrows.meta, shapes.geometric, positioning, calc, scopes,backgrounds, decorations,decorations.pathmorphing,patterns, decorations.pathreplacing}
\pgfplotsset{compat=newest}

\newlength\figureheight
\newlength\figurewidth

\newtheorem{problem}{Problem}
\newtheorem{remark}{Remark}

\DeclareMathOperator{\pre}{Pre}
\DeclareMathOperator{\win}{Win}
\def\binvar{z}

\title{Nonuniform abstractions, refinement and controller synthesis with novel BDD encodings\thanks{This work is supported in part by DARPA grant N66001-14-1-4045, and NSF grants CNS1446298
and ECCS-1553873.}} 

\author{Oscar Lindvall Bulancea\thanks{KTH Royal Institute of Technology, 100 44, Stockholm, Sweden (e-mail: oscarlb@kth.se)}
\and
Petter Nilsson\thanks{California Institute of Technology, Pasadena, CA 91125, USA (e-mail: pettni@caltech.edu)}
\and
Necmiye Ozay\thanks{University of Michigan, Ann Arbor, MI 48109, USA (e-mail: necmiye@umich.edu)}}
\date{}

\begin{document}
\clubpenalty=10000 
\widowpenalty = 10000
\maketitle
\begin{abstract}
This paper presents a control synthesis algorithm for dynamical systems to satisfy specifications given in a fragment of linear temporal logic. It is based on an  abstraction-refinement scheme with nonuniform partitions of the state space. A novel encoding of the resulting transition system is proposed that uses binary decision diagrams for efficiency. We discuss several factors affecting scalability and present some benchmark results demonstrating the effectiveness of the new encodings. These ideas are also being implemented on a publicly available prototype tool, ARCS, that we briefly introduce in the paper.
\end{abstract}

\section{Introduction}

Automatic synthesis of embedded control software that meets its specifications by construction provides a rigorous means for the design of cyber-physical control systems. Abstraction-based techniques, where one creates a finite transition system (FTS) corresponding to the continuous or hybrid system to be controlled, and solves a discrete control synthesis problem, has attracted considerable attention in the past decade \cite{belta2017formal,tabuada2009verification}. 

Various software tools have been developed for correct-by-construction control  synthesis. These tools differ by the class of systems (e.g., discrete-time vs. continuous-time; linear, piecewise affine or nonlinear) or specifications that they can handle (e.g., simple safety or reachability \cite{mazo2010pessoa,rungger2013specification,Rungger:2016:STS:2883817.2883834}; expressive fragments of linear temporal logic \cite{filippidis2016,Wongpiromsarn11}), the abstraction techniques that are used (e.g., uniform grid-based \cite{mazo2010pessoa,Rungger:2016:STS:2883817.2883834}, multi-scale \cite{mouelhi2013cosyma}, or partition-based \cite{filippidis2016}), the way they represent the FTSs internally (symbolic or explicit) and the synthesis techniques implemented.

The key challenge in control synthesis is scalability. Factors affecting scalability include the number of discrete states in the FTS, efficiency of computation of transitions between the discrete states based on continuous dynamics, the representation of the FTS, the complexity of the specification, and the complexity of the resulting controller. For instance, structural properties of dynamics such as linearity or monotonicity \cite{coogan2015efficient} are shown to make the computation of transitions easy. Sparsity of the dynamics has recently been exploited together with binary decision diagrams (BDDs), a compact (i.e., memory-efficient) representation of FTSs, to obtain abstractions efficiently \cite{gruber2017sparsity}. However, it is in general unclear how optimizing different factors individually would affect the efficiency of solving the end-to-end synthesis problem.

 
This paper builds on the abstraction-refinement based incremental synthesis approach by \cite{nilsson2017augmented} that handles a slightly more general class of specifications than most of the earlier tools listed above. In particular, the class includes safety, recurrence and persistence components, while allowing augmented finite transition systems as the discrete model, thereby handling fairness-like assumptions. To mitigate the state-explosion problem, a nonuniform partition of the continuous state space is used. The main contribution of the present paper is a novel BDD encoding of the states that takes into account the topology of the partition and that makes it convenient to add new states in the refinement process while preserving structure. The effectiveness of the new encoding is demonstrated with examples. A prototype tool, ARCS, that implements some of these ideas is also introduced.

\section{Overview}

In this section we formally state the control synthesis problem and give an overview of the solution methodology. 


The first ingredient of the synthesis problem is a dynamical system model
\begin{equation}\label{eq:sys}
x^+ = f(x, u, d), 
\end{equation}
where $x \in 
\mathcal{X}$ is the state, $u\in \mathcal{U}$ is the control input, and $d \in \mathcal{D}$ is the disturbance. The notation $x^+$ either stands for the value of $x$ in the next time step in the discrete-time setting or the derivative in the continuous-time setting. The second ingredient is the specification, which in this paper is restricted to the following fragment of linear temporal logic:
\begin{equation}\label{eq:spec}
	\varphi = \square A \land \lozenge \square B \land \left( \bigwedge_{i \in I} \square \lozenge G^i \right),
\end{equation}
where $A$, $B$, and $G^i$'s are propositions that indicate the membership of the state in a certain subset of the state space $\mathcal{X}$. The specification $\varphi$ roughly mandates that the state trajectory of \eqref{eq:sys} should never leave the states indicated by $A$ (\emph{safety}); that it should eventually reach the states indicated by $B$ and remain there indefinitely (\emph{persistence}); and that it should visit the states indicated by each $G^i$ infinitely often (\emph{recurrence}). In comparison to other fragments of linear temporal logic, formula \eqref{eq:spec} slightly generalizes the guarantee part of a Generalized Reactivity (1) specification \cite{bloem2012synthesis} and is similar to the guarantee part of a Generalized Rabin (1) formula \cite{ehlers2011generalized}. For a more complete treatment of semantics of LTL, we refer the reader to \cite{baier2008principles}. Given these ingredients, the control synthesis problem can be stated as follows.


\begin{problem} Given a system of the form \eqref{eq:sys} and a specification of the form \eqref{eq:spec}, find a \textbf{control policy} $\mu$, possibly dependent on the state history, and a set of initial conditions $\mathcal{X}_0\subset\mathcal{X}$ called the \textbf{winning set}, such that all closed-loop trajectories starting in $\mathcal{X}_0$ satisfy $\varphi$.
\end{problem}
Ideally, we want to solve for the largest winning set $\mathcal{X}_0$ but since this is  in general hard, we opt for trying to incrementally expand it until it is large enough, or a certificate for its maximality is obtained.



Our solution methodology consists of the following steps:
\begin{enumerate}
\item Construction of an abstraction in the form of a (potentially augmented) FTS: This step requires partitioning the state space into cells based on propositions, finding transitions among cells, and, in case of an augmented FTS, finding transient cells for progress groups.\footnote{Augmented finite transition systems (AFTSs) are FTSs extended with progress groups that capture a type of fairness property. In particular, progress groups identify state-action sets such that the system is guaranteed to eventually leave the corresponding state set when the actions from the set are persistently selected. This is useful in abstractions to encode sets of transient states that do not contain any invariant sets. Our results are valid for AFTSs but for simplicity of exposition we focus on FTSs in the rest of the paper. For details of AFTSs, the reader is referred to \cite{nilsson2017augmented}.} 
\item Representing the abstraction using BDDs: This step requires deciding on an encoding of the states and representing transitions and progress groups as a BDD as they are generated in step (1).
\item Solving the discrete synthesis problem on the abstraction via fixed-point algorithms that generate a winning set.
\item Refining the abstraction: If the winning set is not satisfactory, additional cells are added to the partition using the information from the fixed points in step (3). The BDD representation of the abstraction is updated accordingly with new states, transitions, and possibly progress groups.
\item Extracting a controller from the resulting fixed point.
\end{enumerate}

For the first step, we follow the optimization-based procedures in \cite{nilsson2017augmented}. Our tool ARCS currently supports polynomial $f$ for the dynamics, a finite set $\mathcal{U}$ for the inputs, and rectangular sets for $\mathcal{X}$ and $\mathcal{D}$. It can however be extended to any setting where computing or over-approximating reachable sets (required for encoding transitions in the FTS) and certifying transience properties (required for progress groups in an augmented FTS) are possible. This paper is primarily concerned with steps (2) through (5) above; in the rest of the paper we introduce our novel ideas and demonstrate resulting computational gains.

\section{Representing the Abstract System}

As mentioned in earlier sections the continuous system \eqref{eq:sys} is abstracted to an FTS. Formally, an FTS is a tuple $\mathcal{T} = (\mathcal{Q}, \mathcal{U},\rightarrow_\mathcal{T}, \mathcal{L})$, where $\mathcal{Q}$ is a finite set of states, $\mathcal{U}$ is a finite set of inputs, $\rightarrow_\mathcal{T}\subset \mathcal{Q}\times \mathcal{U} \times \mathcal{Q}$ is a transition relation, and $\mathcal{L}$ is a labeling function mapping each state in $\mathcal{Q}$ to a subset of propositions appearing in the formula \eqref{eq:spec}. In order to find a control policy $\mu_\mathcal{T}$ for the FTS $\mathcal{T}$, we need to represent it with a data structure suitable for both storage (memory efficiency) and processing (time efficiency). In what follows, we discuss different representations available in ARCS, their advantages and disadvantages, with a particular focus on a novel BDD encoding. 

\subsection{List Representation}


The perhaps most obvious way to represent an FTS is by encoding states and actions as integers, and transitions as an array 
\begin{equation}
\label{eq:listrep}
\mathcal L = \left[(q_0, u_0, q_0'), (q_1, u_1, q_1'), \ldots, (q_{|\mathcal T|}, u_{|\mathcal T|}, q'_{|\mathcal T|})\right].
\end{equation}

Evidently, this choice requires $\mathcal O(|\mathcal T|)$ memory and standard array operations such as access, insertion, search, and deletion can be done in at most $\mathcal O(|\mathcal T|)$ time, where $|\mathcal{T}|$ is the number of transitions in the system. 

\begin{remark} 
	Eq. \eqref{eq:listrep} can be viewed as a representation of a sparse matrix in coordinate (COO) format, that has non-zero entries $q_k'$ at positions $(q_k, u_k)$. There are several other methods for sparse matrix representations that have different benefits. For instance, the compressed sparse row (CSR) format allows for efficient matrix-vector product computation. \hfill $\blacksquare$
\end{remark}

This representation is very simple and thus easy to implement, and scales linearly in both space and time. Although this might seem like an acceptable complexity, the size of a grid-based abstraction $\mathcal T$ scales at least exponentially with the dimension of the concrete system. To illustrate the potential for improvement consider an $n$-dimensional linear system $\dot x = Ax$: it requires $n^2$ numbers (the entries of $A$) to represent in its canonical ODE form, whereas the size of a finite abstraction based on a list scales exponentially with $n$. This results from the fact that the semantics of an ODE encodes more side information than the semantics of a transition system, thus allowing the former to be more succinct. The idea of this paper is to explore whether information such as geometrical relationships embedded in an ODE can be stored as part of the encoding  of $\mathcal T$ by working with more sophisticated representations. 

\subsection{Binary Decision Diagrams}
{\color{black}In this section, we present an alternative representation of transition systems based on Binary Decision Diagrams (BDDs). We briefly overview how certain operations on transition systems can be performed with BDDs. A key design choice for this type of representations is how to encode the state and action sets as binary variables. As the main contribution of the paper, we present a novel choice for the encodings that attempts to capture underlying geometrical relationships.}

A BDD is a data structure for representing \emph{boolean functions}
\begin{equation}
B : \{0,1\}^n \rightarrow \{0,1\},
\end{equation}
taking binary variables, $\binvar_1$, \dots, $\binvar_n$, defined with an order of evaluation $\binvar_1 < \dots < \binvar_n$. To represent a finite set $C$ with a BDD, one needs an \emph{encoding} $E: C \to \{0,1\}^n$, an injective map from elements of $C$ to truth assignments of the variables $\{\binvar_i\}$, following the order of evaluation.
The boolean function $B_C$ is then said to represent the set $C$ if such an encoding $E$ is defined on all possible elements $c$ and
\begin{equation}\label{eq:BDD repr. def 2}
C = \{c: B_C(E(c)) = 1\},
\end{equation}
i.e. the BDD forms the characteristic function of the set \cite{Bryant1992BDDs}.
For a given encoding such a function can easily be obtained for singletons: if the encoding of $q_k$ is the binary array $E(q_k) = (b_{k,1}, b_{k,2}, \dots, b_{k,n})$, the boolean function for that element can be constructed as
\begin{equation}\label{eq:Element BDD}
B_{q_k}(\binvar_1,\binvar_2,\dots, \binvar_n) = \bigwedge\limits_{i=1}^n \left\{\begin{aligned}
\binvar_i& &b_{k, i} = 1\\
\overline{\binvar}_i& &b_{k,i} = 0
\end{aligned}\right\},
\end{equation}
where $\overline{\binvar}_i$ denotes negation of the variable $\binvar_i$. Then a boolean function for the whole set $C$ can be formed as $B_C = \lor_{k} B_{q_k}$.

To construct BDDs for the elements of a transition mapping $\rightarrow_\mathcal{T}$, an encoding has to be chosen to represent elements $(q_k, u_k, q'_k) \in \mathcal{Q} \times \mathcal{U} \times \mathcal{Q}$. To construct singleton BDDs according to \eqref{eq:Element BDD}, one needs to separate the logical variables for the different parts of the elements, while also separating those used for the initial and final transition states $q_k$ and $q'_k$. Therefore $2n + m$ variables are defined: $\binvar_{q,1},\dots, \binvar_{q,n}$ to represent the set initial states $\mathcal{Q}$, $\binvar_{u,1},\dots, \binvar_{u,m}$ for the action set, and $\binvar_{q',1},\dots,\binvar_{q',n}$ for the final state set $\mathcal{Q}'$. Having defined encodings $E_\mathcal{Q}$ and $E_\mathcal{U}$ for the set of states $\mathcal{Q}$ and set of actions $\mathcal{U}$, an encoding for the transition can be chosen as
\begin{equation}
E_{\mathcal{T}}(q_k, u_k, q'_k) = (E_\mathcal{Q}(q_k), E_\mathcal{U}(u_k), E_\mathcal{Q}(q'_k)).
\end{equation}
With such an encoding, the BDD for one transition $t = (q_k, u_k, q'_k)$ can be constructed as in \eqref{eq:Element BDD}:
\begin{equation}
B_t = B_{q_k} \land B_{u_k} \land B_{q'_k}.
\end{equation}
The BDD representing the entire set $\rightarrow_{\mathcal T}$ is then constructed on disjunctive normal form from the singleton BDDs as
\begin{equation}
B_\mathcal{T} = \bigvee\limits_{i=1}^{|\mathcal{T}|} B_{t_i} =  \bigvee\limits_{i=1}^{|\mathcal{T}|} (B_{q_i} \land B_{u_i} \land B_{q'_i}).
\end{equation}

Working with BDDs as representations of sets, one is limited to the logical operations of boolean functions, among which $\{\land, \lor, \neg, \exists, \forall\}$ are useful. However, using BDDs is analogous to working with the sets themselves, as the basic set operations \{$\cup$, $\cap$, $\setminus$\} have as counterparts logical operations on the corresponding BDDs. It follows from \eqref{eq:BDD repr. def 2} that for sets $C$ and $D$ represented by BDDs $B_C$ and $B_D$:
\begin{align}
B_{C \cup D} &= B_C \lor B_D,\\
B_{C \cap D} &= B_C \land B_D,\\
B_{C \setminus D} &= B_C \land \neg B_D.
\end{align}
In less trivial set definitions, the description might make use of existential and universal quantifiers, $\exists$ and $\forall$. As logical operators, these are defined as acting on a boolean function with one of its variables according to
\begin{align}
\exists_{\binvar_1} B(\binvar_1, \binvar_2) &= B(0, \binvar_2) \lor B(1, \binvar_2) \label{eq:exists def},\\
\forall_{\binvar_1} B(\binvar_1, \binvar_2) &= B(0, \binvar_2) \land B(1, \binvar_2) \label{eq:forall def}.
\end{align}
But as set notation, they are used to signify conditions which have to hold for at least one, or all, elements of a certain set. The BDD equivalent of quantification over elements of a set is to use logical quantification over all variables used in describing the corresponding set BDD. We therefore define the quantifiers $\exists_C$, $\forall_C$ as being
\begin{align}
&\exists_C = \exists_{\binvar_1}\exists_{\binvar_2}\dots \exists_{\binvar_n} \label{eq:exists},\\
&\forall_C = \forall_{\binvar_1}\forall_{\binvar_2}\dots \forall_{\binvar_n} \label{eq:forall},
\end{align}
where the variables $\vec{z} = (\binvar_1, \dots, \binvar_n)$ are used to describe the elements in a set $C$. By expansion of $\eqref{eq:exists}$ and $\eqref{eq:forall}$ using $\eqref{eq:exists def}$ and $\eqref{eq:forall def}$ it can be seen that these operators act on a BDD $B(\vec{z}, \vec{z}')$, where the variables $\vec{z}$ are quantified, by creating new BDDs $(\exists_C B) (\vec{z}')$ and $(\forall_C B) (\vec{z}')$. These new boolean functions have the following properties: ($\exists_C B)(\vec{z}') = 1$ if and only if there exists at least one assignment for $\vec{z}$ such that $B(\vec{z},\vec{z}') = 1$, and, respectively, $(\forall_C B) (\vec{z}') = 1$ if and only if for all assignments to $\vec{z}$, $B(\vec{z}, \vec{z}') = 1$. As assignments are merely the representations of set elements in our definitions, the operators \eqref{eq:exists def}-\eqref{eq:forall def} can be used to construct BDDs that represent sets defined through quantification.

For the operations above, the complexity as functions of the size of BDDs involved is as follows:
\begin{itemize}
\item Conjunction/disjunction of two BDDs $B_1$ and $B_2$ requires $\mathcal{O}(|B_1||B_2|)$ time, producing a BDD with the same bound in size \cite{Bryant1986BDDs}.
\item Negation and assignment of a number of variables in a BDD $B$ requires $\mathcal{O}(|B|)$ time. Negation does not change the BDD size, but the size after variable assignment is bounded by the original BDD size $|B|$ \cite{Bryant1992BDDs}.
\item Following the previous remarks regarding \eqref{eq:exists def} and \eqref{eq:forall def}, quantification of a single variable on a BDD $B$, can be achieved in $\mathcal{O}(|B|^2)$ and results in a size bounded by $\mathcal{O}(|B|^2)$. 
\end{itemize}
Each operation only takes time, and produces a BDD of size, that is polynomial in the input sizes, but successive applications of these operations are required in BDD manipulation. For instance, the quantifications \eqref{eq:exists} and \eqref{eq:forall} have a worst-case complexity that is exponential in the input size. However, these worst-case complexities are seldom encountered in practice.

The BDD data structure is based on a reduced binary tree whose size, i.e. number of nodes, varies not only with the number of elements it represents but also with the encodings and the evaluation order defined for the variables. Choices of evaluation order and encodings are therefore vital when using BDDs and deserves careful consideration, as time and memory used by the logical operations are dependent on the size of the BDD structures involved \cite{Bryant1986BDDs}.

As for the choice of variable ordering, an optimal choice can result in a BDD of size linear in the number of binary variables, and a bad choice can give a size exponential in the number of variables \cite{Bryant1986BDDs}. Finding the optimal variable ordering is a computationally hard problem \cite{Bollig1996} and cannot be solved exactly for any large number of variables in reasonable time, although several heuristics exist (e.g. \cite{Bollig95simulatedannealing}, \cite{SomenziHeuristics}).

The choice of element encoding $E$ involves two aspects: The number of variables used in the encoding, and how each element is mapped. We investigate two kinds of encodings for the states in the abstracted system. One memory-efficient encoding that minimizes the number of variables used, and one encoding that attempts to capture the structure of the partition after iterated refinement.

\paragraph{State Encodings:}
The first type of encoding, which we refer to as the \emph{log encoding}, assumes a numbering of the states from 1 to some number $|\mathcal{Q}|$ and uses these to define the mapping in the form
\begin{equation}\label{eq:log encoding map}
E_\text{log} : \left\{\begin{aligned}
&\mathcal{Q} \rightarrow \{0,1\}^n,\\
&q \mapsto \operatorname{Bin}(q-1),
\end{aligned}
\right.
\end{equation}
where $\operatorname{Bin}(q)$ is the binary representation of the number $q$. When the state space $\mathcal{Q}$ is expanded at refinement, a state with number $k$ is split into two new states. One of these is numbered by $k$ and the other by $|\mathcal{Q}|+1$, after which they are encoded according to \eqref{eq:log encoding map}. The number of variables is also incremented if $|\mathcal{Q}| = 2^n$ before refinement, i.e. all encodings for $n$ variables are used by the present states. With this encoding, the absolute minimum of $\lceil \log_2(|\mathcal{Q}|)\rceil$ variables are used to encode the states. As it simply uses the least amount of variables, it is the encoding to prefer when nothing obvious can be stated about the structure of the problem. This is the standard encoding used in some tools \cite{filippidis2016,gruber2017sparsity}.
\todo[inline]{Describe how new states are encoded}

The novel encoding we propose in this paper---denoted the \emph{split encoding}---is based on the splitting procedure during refinement. As the partition grows increasingly non-uniform with time, with a possibly small area becoming increasingly fine in contrast to others, we believe that an encoding that reflects this structure can lead to computational gains.

\setlength\figurewidth{\columnwidth} 
\setlength\figureheight{0.6\columnwidth} 

\begin{figure}[tb]
	\begin{center}
		\footnotesize
		\includegraphics[width=\columnwidth]{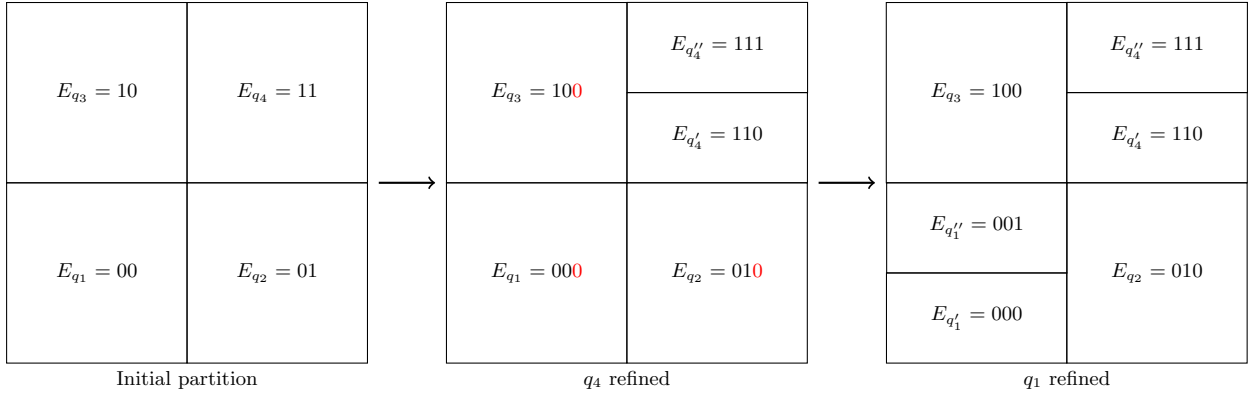}
	\end{center}
	\caption{\small Example of change in split encoding after refinement of initial partition using $k=2$ variables. Refinement of $q_4$ also reaches new largest refinement depth. Bits appended during expansion, but not later assigned, are shown in red.}
	\label{fig:refinement_example}
\end{figure}

\todo[inline]{To Oscar: the description below sounds as if at each refinement step one would add a bit. For example in Fig. 2., if we split $E_{q_1}$ in the next step, is it going to be $(001, 000)$ or $(0001, 0000)$? I thought in the new encoding, the unsplitted cells won't change their encodings.}

Starting with a coarse initial abstraction and an encoding $E$ for the states using $k$ variables, new states created from the splitting procedure have their encoding chosen based on that of its predecessor and refinement depth. The refinement depth is a measure of how many refinements have been performed on the domain the state contains. Every cell resulting from splitting a cell with depth $d$ has a depth of $d+1$, and cells in the initial partition are defined to have depth zero. 
When a state $q$ with depth $d$ is split into two others, $q'$, $q''$, the new states keep the encoding of their predecessor, with a modified bit at position $k+d+1$. This bit is set to 0 for $q'$ and to 1 for $q''$. In the event that the partition reaches a new largest splitting depth, a new variable first has to be created to describe the new states, effectively expanding all encodings by one bit. The default value of the appended bit is chosen as 0. An example of both cases is shown in Fig. \ref{fig:refinement_example}. If the states of the initial partition are given unique initial encodings before refinement, this procedure also results in a unique assignment of state encodings.

To summarize, having a partition with an encoding $E$ and largest refinement depth $D$; the state $q$ with a refinement depth $d$ and encoding $E(q) = (b_{q,1},\dots, b_{q,k},\dots,b_{q,k+D})$, is split into two states labeled $q'$ and $q''$. The encodings of the new states are chosen as
\begin{align}
&E(q') = E(q)\big|_{b_{k+d+1} = 0},\label{eq:split enc 0}\\
&E(q'') = E(q)\big|_{b_{k+d+1} = 1}\label{eq:split enc 1}.
\end{align}
But if $d = D$, then all encodings are first expanded by a 0 bit, i.e.
\begin{equation}
E(q) = (E(q), \thinspace 0) \quad \forall q \in \mathcal{Q},
\end{equation}
before applying \eqref{eq:split enc 0} and \eqref{eq:split enc 1} with the newly expanded encodings.\footnote{Appending the encodings  with $0$ to have all states represented with the same number of bits is just for coding convenience. One can alternatively change the co-domain of the function $E$ to $\cup_{i=k}^n \{0,1\}^i$.}

With this choice of encoding, neighboring cells have similar encodings and our hypothesis is that such similarity admits a compact BDD representation by capturing the geometry of the underlying vector fields. Smaller BDDs typically result in computational gains in the synthesis step, an effect we see in simulations. However, the number of binary variables is generally larger than with the log encoding so the theoretical worst-case complexity is higher.


\section{Solving Synthesis Problems}

The general way of solving a finite LTL synthesis problem involves translating the LTL specification to a Rabin automaton and computing fixed points on the product of the transition system and the Rabin automaton. Certain LTL fragments do however admit winning sets defined as fixed points on the transition system, which avoids the potentially expensive construction of the product system. This is the case for the GR(1) fragment, as well as the fragment \eqref{eq:spec} considered in this paper. In the following, we present the fixed-point mappings associated with \eqref{eq:spec} and how they can be evaluated symbolically when sets are represented as BDDs.

The fundamental component of the fixed-point mappings is the backwards controlled reachability operator $\pre_{\sharp_1, \sharp_2}$ defined as follows:
\begin{equation}\label{eq:pre def}
	\pre_{\sharp_1, \sharp_2}(X) \hspace{-1mm} = \hspace{-1mm} \left\{ q \hspace{-0.5mm} : \hspace{-0.5mm} (\sharp_1  u \hspace{-0.5mm} \in \mathcal U), (\sharp_2 (q, u, q') \in \rightarrow_\mathcal{T}), q' \hspace{-0.5mm}  \in \hspace{-0.5mm} X \right\}.
\end{equation}
Here, $\sharp_1$ and $\sharp_2$ is either $\exists$ or $\forall$ and reflect the controllability assumptions: $\pre_{\exists, \forall}$ corresponds to $u$ being controllable and nondeterminism uncontrollable, whereas $\pre_{\forall, \exists}$ corresponds to uncontrollable $u$ but controllable nondeterminism.

\paragraph{Computing $\pre$ with list representation:} For synthesis algorithms the fundamental operator is the $\pre_{\exists, \forall}(X)$ operator, which can be computed as follows:
\begin{enumerate}
\item Find the set $C = \pre_{\exists, \exists}(X)$ of all $q$ such that there exists $(q,u,q') \in \rightarrow_{\mathcal T}$ for $q' \in X$.
\item For each $q \in C$, for each action $u$, find the set $C_{q,u} = \{ q' : \exists (q, u, q') \in \rightarrow_{\mathcal T} \}$.
\item Now for all $q$, $q \in \pre_{\exists, \forall}(X)$ if and only if $q \in C$ and for some $u$, $C_{q,u} \neq \emptyset$ and $C_{q,u} \subset X$.
\end{enumerate}
The procedure can be slightly modified to account for other combinations of quantifiers (i.e. $\forall, \forall$). The first step can be done via one traversal of $\mathcal L$, and the second via $|C| |\mathcal U|$ traversals. Thus the complexity for computing $\pre$ is upper bounded by $\mathcal O \left( |\pre_{\exists, \exists}(X)| |\mathcal U| |\mathcal T| \right)$. However, since the same sets $C_{q,u}$ are typically computed many times when evaluating a fixed point, step 2 can be discounted across multiple $\pre$ operations  by storing the sets $C_{q,u}$, which improves the time complexity of a single $\pre$ computation towards $\mathcal O(|\mathcal T|)$ at the expense of a larger memory footprint.

\paragraph{Computing $\pre$ with BDD representation:} When the set of final states $X$ is represented as a BDD $B_X$ using an encoding $E$, the set $\pre_{\sharp_1, \sharp_2}(X)$ can be represented as the binary mapping
\begin{equation}\label{eq:pre BDD}
	B_{\pre_{\sharp_1, \sharp_2}(X)} = \left\{\begin{aligned}
    &{\sharp_1}_{\mathcal{U}}{\exists}_{\mathcal{Q}'}(B_{\mathcal T} \land B_X), &\sharp_2 = \exists,\\
    &{\sharp_1}_{\mathcal{U}}{\forall}_{\mathcal{Q}'}(\neg B_{\mathcal T} \lor B_X), &\sharp_2 = \forall, 
    \end{aligned}\right.
\end{equation}
which can be computed symbolically from $B_X$ via quantifier elimination.\footnote{This is kept simple for the reader, however one has to be careful when performing negations, as the domain of possible assignments could be larger than the domain of elements. As such, one would need to modify \eqref{eq:pre BDD}, to $B_\mathcal{Q} \land \sharp_\mathcal{U}(\forall_{\mathcal{Q}'}(\neg B_\mathcal{T} \lor B_X))$ if $\sharp_2 = \forall$, further replace $\exists_\mathcal{U}(\dots)$ with $\exists_\mathcal{U}(B_\mathcal{U} \land \dots)$ if also $\sharp_1 = \exists$, and if $(\sharp_1, \sharp_2) = (\forall, \exists)$ replace $\forall_\mathcal{U}(\dots)$ with $\forall_\mathcal{U}(\neg B_\mathcal{U} \lor \dots)$, to not include assignments corresponding to non-existent states.}
The runtime of this operation ultimately depends on the sizes of intermediate results, but considering the complexity and worst-case size result of each operation involved, an upper bound can be obtained as $\mathcal{O}((|B_X||B_\mathcal{T}|)^{2^{n_u + n_q}})$, when using $n_u$ action variables and $n_q$ end state variables. The complexity can also be expressed solely in the number of variables used, if one assumes the maximum size of each intermediate BDD. Then the theoretically largest possible runtime is that of the $2^{n_q-1}$ conjunctions when expanding ${\sharp_2}_{\mathcal{Q}'}$, acting on BDDs having assigned all but $n_q + n_u$ variables, thus having complexity $\mathcal{O}(2^{3n_q + 2n_u})$.
\todo[inline]{Stated complexity on both input size and variable forms. Useful to have both?}

Equipped with the $\pre$ operator, we give fixed-point characterizations for the winning set of \eqref{eq:spec}. We borrow notation from $\mu$-calculus for succinct expression of fixed points. Let $\kappa : 2^{\mathcal{Q}} \rightarrow 2^{\mathcal{Q}}$ be a mapping that is monotone with respect to set inclusion, i.e., $V \subset W \implies \kappa(V) \subset \kappa(W)$. Then the \emph{greatest fixed point of} $\kappa$, written $\nu V \; \kappa(V)$, is the value after convergence of the set sequence
\begin{equation}
\label{eq:largefp}
        V_0 = \mathcal{Q}, \quad V_{k+1} = \kappa(V_k).
\end{equation}
Correspondingly, the \emph{smallest fixed point of} $\kappa$, written $\mu V \; \kappa(V)$, is the value after convergence of
\begin{equation}
\label{eq:smallfp}
        V_0 = \emptyset, \quad V_{k+1} = \kappa(V_k).
\end{equation}
Due to monotonicity and finiteness of $\mathcal{Q}$, both these sequences converge in a finite number of steps. With this notation, the winning set of \eqref{eq:spec} is as follows:
\begin{equation}
\label{eq:synth_fp}
\begin{aligned}
    \win_{\exists, \forall} & \left( \varphi \right)  = \mu V_2 \; \nu V_1 \; \bigcap_{i \in I} \mu V_0 \; \pre_{\exists, \forall}(V_2) \\ & \hspace{-3mm} \cup \left( B \cap G^i \cap \pre_{\exists, \forall}(V_1) \right) \cup \left( B  \cap \pre_{\exists, \forall}(V_0) \right).
\end{aligned}
\end{equation}

\todo[inline]{The advantage of the fragment in \eqref{eq:spec}: fixed points directly definable on the transition system, similar to GR(1), rather than automaton construction, which might be costly.}
\todo[inline]{Focus on fixed points and how they are implemented with BDDs?}
\todo[inline]{Refinement, how it changes the encoding}


\subsection{Specification-Guided Abstraction Refinement}

In the event that the winning set $\win_{\exists, \forall}(\varphi)$ computed via \eqref{eq:synth_fp} is empty, or otherwise not satisfactory (e.g., it does not cover an expected initial condition), the abstraction can be refined in an attempt to extract more information about the underlying concrete system. Instead of doing this blindly, we select refinement regions guided by the internals of the fixed point computation \eqref{eq:synth_fp}. Loosely speaking, for a greatest fixed point \eqref{eq:largefp} we perform refinement in the set $\pre_{\exists, \exists}(V_\infty) \setminus V_\infty$ just outside the fixed point $V_\infty$, with the hope that the greatest fixed point will be enlarged in the refined system. For a smallest fixed point, refinement is instead done in $V_1 \setminus V_\infty$, where $V_k$ is the $k$'th iteration of \eqref{eq:smallfp}. These ideas are illustrated in Fig. \ref{fig:refinement}. For multi-level fixed points such as \eqref{eq:synth_fp} we select the refinement regions as the union of the refinement regions corresponding to lower-level fixed points. For more details see \cite{nilsson2017augmented}.

\definecolor{mygreen}{rgb}{0.0, 0.6, 0.0}

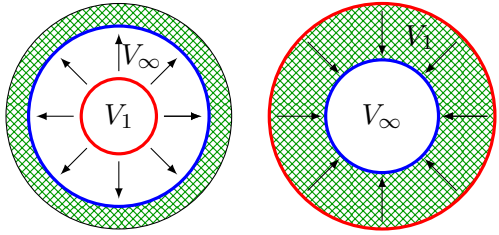
\begin{figure}
  \begin{center}
  \begin{tikzpicture}
      \draw[pattern=crosshatch, pattern color=mygreen, draw=none] (3.5,0) circle (1.5);
      \draw[red, very thick] (3.5,0) circle (1.5);
      \draw[blue, fill=white, very thick] (3.5,0) circle (0.75);

      \draw[pattern=crosshatch, pattern color=mygreen] (0,0) circle (1.5);
      \draw[fill=white] (0,0) circle (1.2);
      \draw[fill=white, draw=blue, very thick] (0,0) circle (1.2);
      \draw[red, very thick] (0,0) circle (0.5);

      \foreach \s in {0,45,...,360} {
          \draw[-latex] (\s:0.6) -- ++ (\s:0.5);
          \draw[-latex] ($(3.5,0) + (\s:1.4)$) -- ++ (\s:-0.6);
      }
      
      \node at (0,0) {$V_1$};
      \node at (0.3,0.8) {$V_\infty$};
      \node at (3.5,0) {$V_\infty$};
      \node at (4,1.05) {$V_1$};
  \end{tikzpicture}
  \end{center}
  \caption{\small Illustration of fixed point-based refinement; the initial set $V_1$ is in red and the final fixed point $V_\infty$ in blue. Refinement is done in the green shaded regions. Left: for a smallest fixed point \eqref{eq:smallfp} the set sequence $\{ V_k \}$ is monotonically expanding (w.r.t set inclusion), and refinement is done in $\pre_{\exists, \exists}(V_\infty) \setminus V_\infty$. Right: for a greatest fixed point \eqref{eq:largefp} the set sequence $\{ V_k \}$ is monotonically contracting, and refinement is done in $V_1 \setminus V_\infty$.}
  \label{fig:refinement}
\end{figure}

\subsection{Controller Extraction}

In addition to computing the winning set, in practice also a controller that enforces the specification inside the winning set is required. Fundamentally, such a controller can be extracted by saving the set of $u$'s satisfying the quantification in low-level calls to $\pre$ in \eqref{eq:pre def}, and storing these $u$'s in a memory hierarchy whose structure depends on the type of fixed point. For instance, invariance or reachability controllers can be memoryless (i.e. state feedback), but a recurrence controller must maintain an internal memory state that switches between different reachability objectives. {\color{black}Thus certain specifications trivialize controller extraction and as a result this issue is not discussed in detail for tools that handle only invariance specifications. However, control extraction adds to the overall complexity when specifications of more general form such as \eqref{eq:spec} are considered, or when AFTSs are used.}

\todo[inline]{Oscar: I could add controller extraction for BDDs, Needs to be done in detail for the lowest level (made together with Pre computation), but no different for the rest except replacement of set operations with BDD logic operations (but perhaps shouldn't guarantee this as I haven't implemented it correctly yet).}

\todo[inline]{Maybe we should mention here that for other tools that only do reachability and safety, controller extraction is trivial, controllers are memoryless. For richer specs as ours, one needs to be careful.}

\section{Results and Comparisons}
{\color{black}In this section, we present results comparing different representations of transition systems. Our toolbox ARCS, available at \url{https://github.com/pettni/abstr-refinement}, implements the examples in this section. ARCS has a MATLAB front-end for handling continuous dynamics, computation of transitions, and list representations,  and a C back-end for BDD operations using the CUDD library \cite{cudd}.}

As benchmarks we consider hydronic radiant systems for buildings, in which chilled water is run through concrete slabs to regulate the temperature of the rooms to which they are connected. The systems are controlled by turning on/off flow to any one slab, thus changing the temperature of zone $i$ according to the heating dynamics
\begin{equation}
c_i\dot{T}_i = \sum_{j\neq i}\frac{1}{R_{ij}}(T_i - T_j) + k_i,
\end{equation}
where the sum is taken over all temperature zones in the system, including other rooms, slabs, supply water sources, and the outside (the last two are assumed to have constant temperature). Thermal capacitances $c_i$, thermal resistances $R_{ij}=R_{ji}$ and nominal heat gains $k_i$ are determined by sets of parameters that together define room types. In this article we use parameters for the outside and water temperature and two room types as defined in \cite{nilsson2017augmented}, slightly extending the framework to take into account adjacency with multiple rooms of different types, and multiple slabs in different configurations. A setup with $n_r$ rooms and $n_s$ slabs results in a dynamical system with $n_r+n_s$ continuous states (room and slab temperatures), and $2^{n_s}$ discrete control inputs (water flow through each slab turned on or off).

We construct a collection of benchmark systems and perform three types of run time tests:
\begin{enumerate}
	\item Synthesis on systems of varying size and topology,
	\item Synthesis on the same system but at different levels of refinement,
    \item An end-to-end synthesis-refinement procedure on one system.
\end{enumerate}
For these tests a persistence specification $\varphi = \lozenge \square B$ is considered, with $B$ being the proposition corresponding to all rooms and slabs having temperatures in $[22,25]^\circ C$ and $[21 , 27]^\circ C$ respectively, in a total domain of $([20, 28] ^\circ C)^{n_r + n_s}$.

\begin{figure}[tb]
	\begin{center}
		\footnotesize
		\includegraphics[width=\columnwidth]{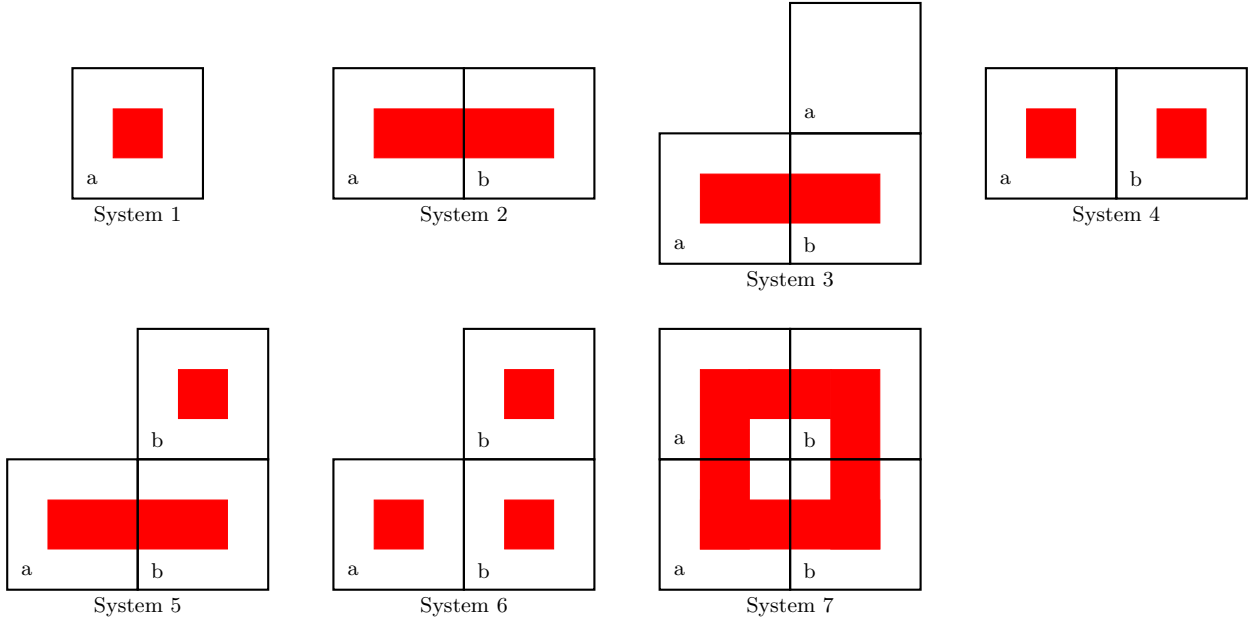}
	\end{center}
	\caption{\small Test configurations based on different building layouts. Rooms of different types (a or b) having different heating dynamics, controlled by cooling slabs (red).}
	\label{fig:configs}
\end{figure}

In the first test, different configurations of adjacent rooms and connecting slabs, as shown in Fig. \ref{fig:configs}, are used. After refining the abstraction of each system 2000 times, resulting in roughly 2000 states in each abstraction, a final synthesis is performed and timed. The final synthesis is done using the BDD representation with the suggested log and split encoding, both as is and after reordering of the BDD variables using the simulated annealing algorithm implemented in CUDD. Fig. \ref{fig:figure1} shows the resulting run times for each system and representation scenario. It is worth noting that for all systems except 3, which is special in the sense that one room lacks a controller, the order from slowest to fastest run times is consistently measured as log, reordered log, split and reordered split encodings.

\setlength\figurewidth{0.6\columnwidth} 
\setlength\figureheight{0.3\columnwidth}

\begin{figure}[tb]
	\begin{center}
		\footnotesize
		\begin{tikzpicture}
    \begin{semilogyaxis}[
        width = \figurewidth,
        height = \figureheight,
        legend style={at={(0.95,0.025)},anchor=south east, font=\tiny},
        legend cell align=left,
        xlabel = {Test number},
        ylabel = {Run time [s]}]

        \addplot[blue, line width=1pt] table[x=num, y=log] {figures/data/run_time_data_by_id.csv};
        \addlegendentry{Log};

        \addplot[blue!50, dashed, line width=1pt] table[x=num, y=re-log] {figures/data/run_time_data_by_id.csv};
        \addlegendentry{Log reordered};

        \addplot[red, line width=1pt] table[x=num, y=split] {figures/data/run_time_data_by_id.csv};
        \addlegendentry{Split};

        \addplot[red!50, dashed, line width=1pt] table[x=num, y=re-split] {figures/data/run_time_data_by_id.csv};
        \addlegendentry{Split reordered};
    \end{semilogyaxis}
\end{tikzpicture}
	\end{center}
	\caption{\small Comparison of stand-alone synthesis run times using the different versions of BDD encoding, with and without reordering.}
	\label{fig:figure1}
\end{figure}
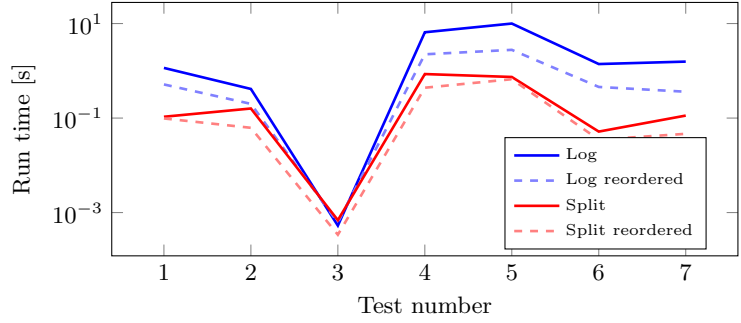

\begin{figure}[tb]
	\begin{center}
		\footnotesize
        \begin{tikzpicture}
    \begin{loglogaxis}[
        width = \figurewidth,
        height = \figureheight,
        ymax=10000,
        legend style={at={(0.05,0.95)},anchor=north west, font=\tiny},
        legend cell align=left,
        xlabel = {Number of transitions},
        ylabel = {Run time [s]},
        grid=major,
        major grid style={black!30}]

        \addplot[blue, line width=1pt] table[x=trans, y=log] {figures/data/complexity_data.csv};
        \addlegendentry{Log};

        \addplot[blue!50, dashed, line width=1pt] table[x=trans, y=re-log] {figures/data/complexity_data.csv};
        \addlegendentry{Log reordered};

        \addplot[red, line width=1pt] table[x=trans, y=split] {figures/data/complexity_data.csv};
        \addlegendentry{Split};

        \addplot[red!50, dashed, line width=1pt] table[x=trans, y=re-split] {figures/data/complexity_data.csv};
        \addlegendentry{Split reordered};
        
        \addplot[green, line width=1pt] table[x=trans, y=list] {figures/data/complexity_data.csv};
        \addlegendentry{List rep.};
    \end{loglogaxis}
\end{tikzpicture}
	\end{center}
    \small
	\caption{\small Comparison of stand-alone synthesis run times for System 2 (Fig. \ref{fig:configs}) using the different representations and BDD encodings, with and without reordering, for variable level of refinement.}
	\label{fig:complexity}
\end{figure}
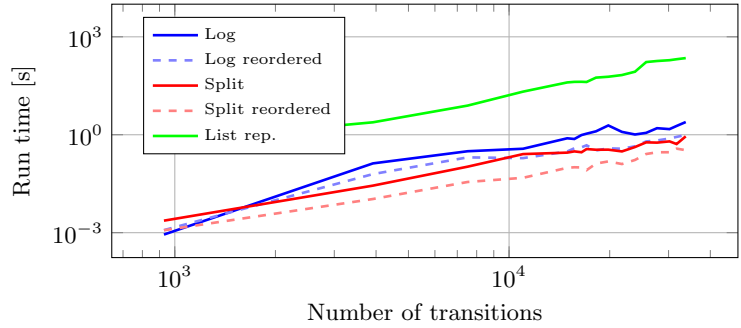

In the second test, System 2 in Fig. \ref{fig:configs} is used in measuring the synthesis time at different levels of refinement. An abstraction is obtained after a number of synthesis-refinement iterations between 500 and 4000, and the run time of synthesis on that abstraction is measured. In Fig. \ref{fig:complexity} the run time of synthesis is plotted against the number of transitions present in the abstraction. Just as in the first test, the split encoding is seen to have lower run time than the log encoding for both the ordered and unordered case. It is also interesting to note how for this test, the slope of the reordered split encoding graph in this loglog-plot is close to linear and less than that of the list encoding toward the larger number of transitions.

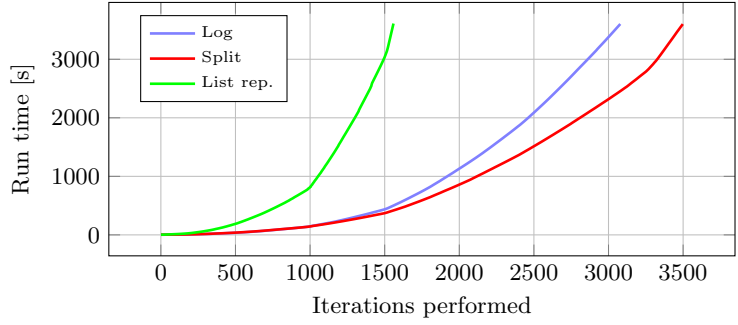
\begin{figure}[tb]
	\begin{center}
		\footnotesize
		\begin{tikzpicture}
    \begin{axis}[
        /pgf/number format/.cd, 1000 sep={},
        width = \figurewidth,
        height = \figureheight,
        legend style={at={(0.05,0.95)},anchor=north west, font=\tiny},
        legend cell align=left,
        xlabel = {Iterations performed},
        ylabel = {Run time [s]},
        grid=major]

        \addplot[blue!50, line width=1pt] table[y=time, x=iter] {figures/data/end_to_end_log.txt};
        \addlegendentry{Log};

        \addplot[red, line width=1pt] table[y=time, x=iter] {figures/data/end_to_end_split.txt};
        \addlegendentry{Split};

        \addplot[green, line width=1pt] table[y=time, x=iter] {figures/data/end_to_end_sparse.txt};
        \addlegendentry{List rep.};
    \end{axis}
\end{tikzpicture}
	\end{center}
	\caption{\small End-to-end synthesis-refinement performance test for System 2 (figure \ref{fig:configs}), letting the synthesis-refinement algorithm run for one hour, using the different representations and BDD encodings.}
	\label{fig:end_to_end}
\end{figure}

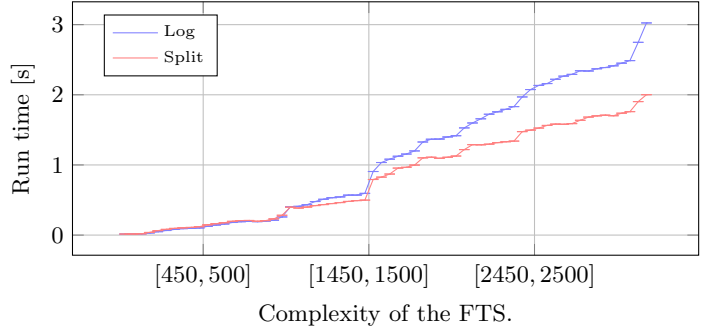
\begin{figure}[tb]
	\begin{center}
		\begin{tikzpicture}
\def\intervSize{50}
\footnotesize
\begin{axis}[
  /pgf/number format/.cd, 1000 sep={},
  xtick={500, 1500, 2500},
  xticklabel=
    {$[\pgfmathparse{\tick-\intervSize}\pgfmathprintnumber{\pgfmathresult},%
	   \pgfmathparse{\tick}\pgfmathprintnumber{\pgfmathresult}]$},
  xlabel = {Complexity of the FTS.},
  ylabel = {Run time [s]},
  legend style={at={(0.05,0.95)},anchor=north west, font=\tiny},
  legend cell align=left,
  grid=major,
  width=\figurewidth,
  height=\figureheight
]
\addplot+[blue!50, mark=none, error bars/.cd, y dir=both] table[x=iters, y=mean, y error=var] {figures/data/end_to_end_histo_log.txt};
\addlegendentry{Log}

\addplot+[red!50, mark=none, error bars/.cd, y dir=both] table[x=iters, y=mean, y error=var] {figures/data/end_to_end_histo_split.txt};
\addlegendentry{Split}
\end{axis}
\end{tikzpicture}
	\end{center}
    \caption{\small Runtime of synthesis for System 2 with FTSs and an increasing number of states generated during refinement steps for the log and split encoding. Mean and variance of the overall timing is reported over intervals of 50 steps.}
    \label{fig:end_to_end_histo}
\end{figure}

\begin{figure}[tb]
	\begin{center}
		\footnotesize
\begin{tikzpicture}






	\begin{axis}[
        /pgf/number format/.cd, 1000 sep={},
        width = \figurewidth,
        height = \figureheight,
        legend style={at={(0.95,0.95)},anchor=north east, font=\tiny},
        legend cell align=left,
        xlabel = {Iterations performed},
        ylabel = {Memory per variable [MB]},
        grid=major]

        \addplot[blue!50, line width=1pt] table[y expr={\thisrowno{3}/(2*\thisrowno{4}+\thisrowno{5})/1024^2}, x=iters] {figures/data/mem_log.txt};
        \addlegendentry{Log};

        \addplot[red, line width=1pt] table[y expr={\thisrowno{3}/(2*\thisrowno{4}+\thisrowno{5})/1024^2}, x=iters] {figures/data/mem_split.txt};
        \addlegendentry{Split};
    \end{axis}
\end{tikzpicture}
	\end{center}
    \caption{\small Memory used per binary variable as a function of refinement steps for System 2.}
    \label{fig:memory}
\end{figure}
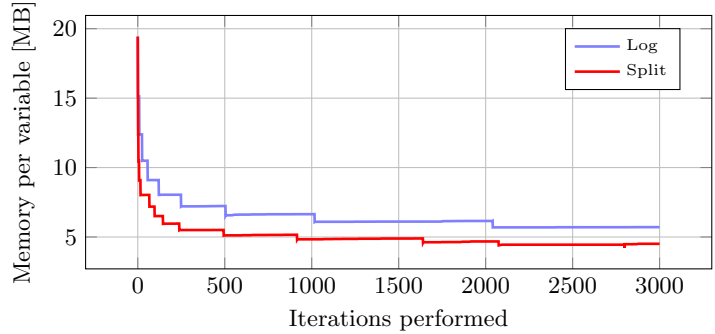

As a final test, the synthesis-refinement procedure is ran on System 2 for one hour using the list representation, log and split encoding, and cumulative time together with runtime of each synthesis-refinement iteration is measured. In Fig. \ref{fig:end_to_end}, the cumulative time is plotted against the total number of iterations achieved, and in Fig. \ref{fig:end_to_end_histo}, the runtime of the iterations are shown as averages over intervals of 50 iterations. As seen in Fig. \ref{fig:end_to_end}, the split encoding manages a few hundred more iterations in total than the log encoding, especially during the time the abstraction is more refined and transition relations are more complex. And in Fig. \ref{fig:end_to_end_histo}, the average iteration time for log encoding grows at a faster rate than for the split encoding, when the partition is increasingly refined. {\color{black}As a side note, overall timing for each of the methods can be improved by warm-starting synthesis at each refinement step (see \cite{nilsson2017augmented}) but this is not currently implemented in ARCS, and will not affect the comparative results significantly. Finally, we also note that in terms of memory requirements, the number of nodes allocated by CUDD to represent the BDDs for split encoding is approximately twice as much as that for the log encoding, with $2\%$ to $10\%$ extra memory usage at iteration steps 2000 and 3000, respectively. This extra memory usage is not surprising given that the number of binary variables in split encoding is more than the minimal number achieved by the log encoding. On the other hand, the split encoding achieves a better compression in terms of memory used per number of binary variables as shown in Fig. \ref{fig:memory}. Moreover, this redundancy in representation seems to improve computation times significantly.}

Finally, we point out how non-uniformly the partition of the state space evolves through the abstraction-refinement process. Figure \ref{fig:init_partition} shows an initial partition of the domain for System 1 and the partition after 200 refinement steps for the same system. It is hard to capture the structure in the refined partition when starting with a log encoding of the initial state and appending one state to the end of the list of states at every refinement step, where the list is eventually encoded via \eqref{eq:log encoding map}. On the other hand, the split encoding is designed to capture the topological relations in the partition naturally. 

\setlength\figurewidth{0.42\columnwidth} 
\setlength\figureheight{0.3\columnwidth}

\begin{figure}[tb]
	\begin{center}
		\footnotesize
		\begin{tikzpicture}
    \begin{axis}
    [enlargelimits=false,
    axis on top,
    width=\figurewidth,
    height=\figureheight,
    xlabel = Slab Temp (\si{\celsius}),
    ylabel = Room Temp (\si{\celsius})]
       \addplot graphics [xmin=20,xmax=28,ymin=20,ymax=28] {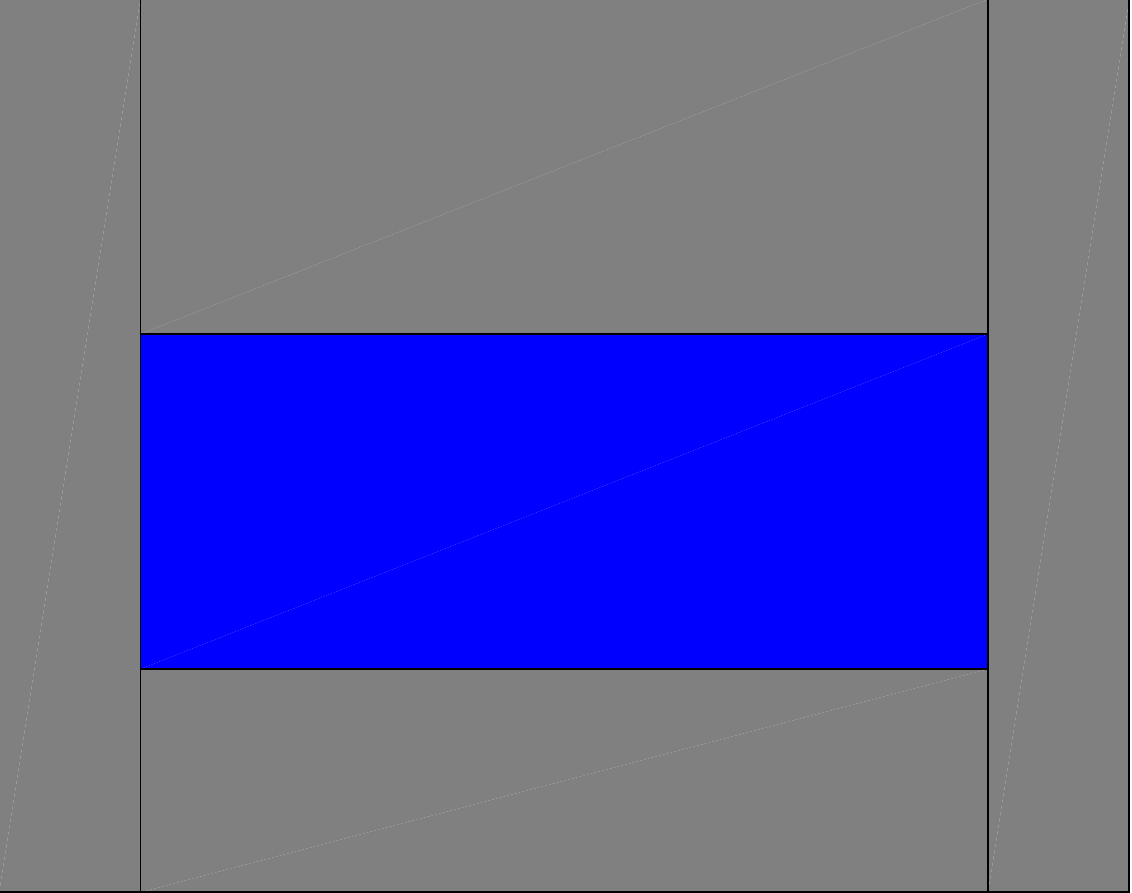};
       
    \end{axis}
\end{tikzpicture} ~ \begin{tikzpicture}
    \begin{axis}
    [enlargelimits=false,
    axis on top,
    width=\figurewidth,
    height=\figureheight,
    xlabel = Slab Temp (\si{\celsius}),
    ylabel = Room Temp (\si{\celsius}),
    legend style={at={(1.05,0.95)},anchor=north west, font=\tiny},
    legend cell align=left]
    	\addlegendimage{area legend, fill=blue};
        \addlegendimage{area legend, fill=green};
       \addplot graphics [xmin=20,xmax=28,ymin=20,ymax=28] {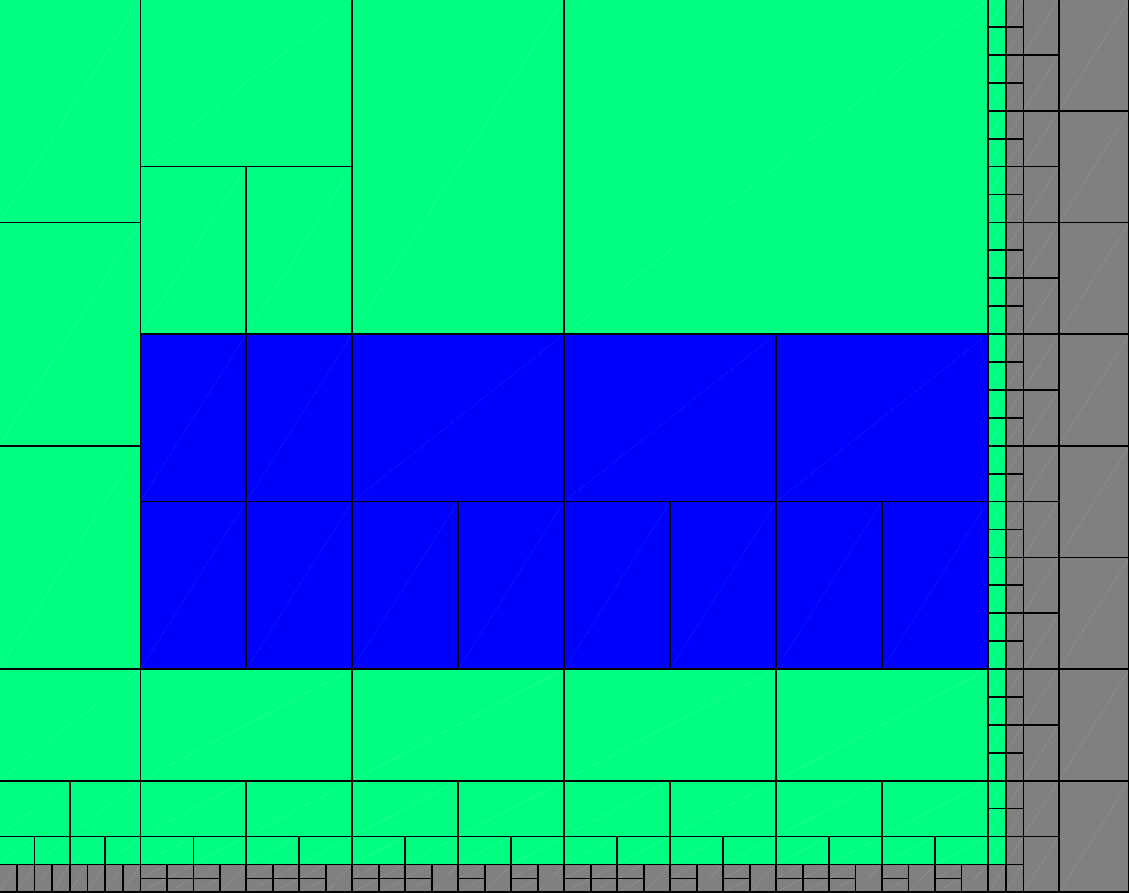};

       \addlegendentry{Goal set};       
       \addlegendentry{Win set};
    \end{axis}
\end{tikzpicture}
	\end{center}
    \caption{Left: Initial partition of System 1 showing domain and goal set. Right: Partition after 200 abstraction-refinement iterations, showing goal and calculated winning set.}
    \label{fig:init_partition}
\end{figure}

\section{Conclusions}

In this paper, we presented an abstraction-refinement based controller synthesis framework and, specifically, discussed several ways of representing the transition systems resulting from abstractions. We proposed a novel BDD-based encoding, namely \emph{split encoding}, of the states of the transition system that takes into account the geometry of the underlying continuous states and how they evolve with refinement. A comparative study of various representations shows the effectiveness of the new encoding. The presented ideas are implemented in a toolbox, ARCS, which is made publicly available. 

\bibliographystyle{abbrv}

\end{document}